\documentstyle[11pt,aasms]{article}
\input psfig.tex

\newcommand\xvec{{\bf x}}

\newbox\grsign \setbox\grsign=\hbox{$>$} \newdimen\grdimen \grdimen=\ht\grsign
\newbox\simlessbox \newbox\simgreatbox
\setbox\simgreatbox=\hbox{\raise.5ex\hbox{$>$}\llap
     {\lower.5ex\hbox{$\sim$}}}\ht1=\grdimen\dp1=0pt
\setbox\simlessbox=\hbox{\raise.5ex\hbox{$<$}\llap
     {\lower.5ex\hbox{$\sim$}}}\ht2=\grdimen\dp2=0pt

\begin{document}

\title{CONSTRAINTS ON THE EFFECTS OF LOCALLY-BIASED GALAXY FORMATION}

\author{Robert J. Scherrer}
\affil{NASA/Fermilab
Astrophysics Center, Fermi National
Accelerator Laboratory, P.O. Box 500,
Batavia, IL  60510, and Department of Physics and Department of Astronomy,
Ohio State University, Columbus, OH 43210}

\author{David H. Weinberg}
\affil{Department of Astronomy, Ohio State University, Columbus, OH 43210}

\begin{abstract}
While it is well-known that ``biased galaxy formation" can 
increase the strength of galaxy clustering, it is less clear 
whether straightforward biasing schemes can change the {\it shape}
of the galaxy correlation function on large scales.
Here we consider ``local" biasing models, in which the galaxy
density field $\delta_g$ at a point {\bf x} is a function 
of the matter density field $\delta$ at that point: 
$\delta_g = f(\delta)$.  We consider both deterministic
biasing, where $f$ is simply a function, and stochastic
biasing, in which the galaxy density $\delta_g$ is a random
variable whose distribution depends on the matter density:
$\delta_g=X(\delta)$.  
We show that even when this mapping is performed on a highly
nonlinear density field with a hierarchical correlation
structure, the correlation function $\xi$ is
simply scaled up by a constant,
as long as $\xi \ll 1$.  In stochastic biasing models,
the galaxy autocorrelation function
behaves exactly as in deterministic models, with
$\overline{X}(\delta)$ (the mean value of $X$ for a given value of $\delta$)
taking the role of the deterministic bias function.  We extend our results
to the power spectrum $P(k)$, showing that for sufficiently small $k$,
the effect of local biasing is equivalent to the multiplication of $P(k)$ by
a constant, with the addition of a constant term.
If a cosmological model predicts a large-scale mass correlation
function in conflict with the shape of the observed galaxy
correlation function, then the model cannot be rescued by appealing to 
a complicated but local relation between galaxies and mass.

\end{abstract}


\vfill
\eject

\section{Introduction}

If galaxies form with greater efficiency (per unit mass) in high
density regions, then their clustering
can be amplified with respect to that of the
underlying mass distribution (\cite{kaiser84}).  
This amplification is often summarized in terms of a ``bias factor'' $b$,
where $b^2 = \xi_g(r)/\xi(r)$ is the ratio of the galaxy autocorrelation
function to the mass autocorrelation function.
Biased galaxy formation
plays a crucial role in cosmological scenarios that
assume a critical density ($\Omega=1$) universe, since these
models predict excessively high velocity dispersions 
in galaxy groups and clusters unless the amplitude of 
mass correlations is lower than the observed amplitude of galaxy correlations
(\cite{davis85}; \cite{bardeen86}).
At first glance, it appears obvious that bias can alter the {\it shape}
of the autocorrelation function in addition to changing the amplitude, 
since one can simply write
the bias factor $b$ as a bias function $b(r)$.
However, a physical theory of biased galaxy formation cannot specify
$b(r)$ directly, it can only specify how the efficiency of galaxy
formation depends on environment --- $b(r)$ is an output of such
a theory, not an input.  
For example,
the widely examined, ``high peak'' model of galaxy formation
predicts a scale-independent
bias factor, at least in the linear regime (\cite{bardeen86}).

The possibility of scale-dependent bias became a serious issue
once it was shown that the shape of the galaxy autocorrelation
function differed from the shape predicted by the ``standard'' cold
dark matter (CDM) model, on large scales close to the linear regime
(\cite{maddox90}).  With scale-dependent bias,
one could in principle resolve
this discrepancy by appealing to the complex astrophysics of galaxy
formation instead of altering the CDM
model's fundamental cosmological assumptions
(e.g., the value of $\Omega$).  However, the specific schemes
that have been proposed to achieve the
requisite scale-dependence are all {\it non-local};
the efficiency of galaxy formation is directly modulated
in a coherent fashion over large scales (\cite{babul91}; \cite{bower93}).
While this sort of coherent modulation is physically possible,
it seems {\it a priori} less natural than models in which the
efficiency of galaxy formation depends only on properties of the
local environment.  Weinberg (1995) and Mann, Peacock, \& Heavens (1997)
applied a wide range of {\it local} biasing schemes to cosmological
N-body simulations, and they found that these schemes did not change
the shape of the galaxy autocorrelation function or 
of its Fourier transform, the
power spectrum, on large scales, though they did alter the shape
in the nonlinear regime.

Is non-locality essential to producing scale-dependent bias on large scales?
In this paper we address this question analytically, extending results from
earlier work.  Coles (1993) showed that an arbitrary
local bias applied to a Gaussian density field 
amplifies (or depresses) the autocorrelation function by a constant
multiplicative factor.  His argument works for Gaussian fields
even when the rms fluctuations are nonlinear, but
in the real universe the nonlinear density field cannot be Gaussian
because densities cannot be negative.
In practice, the efficiency of galaxy formation may depend on the mass
density averaged over some fairly small, nonlinear scale, and there
will almost certainly be scatter about the mean relation between
galaxy and mass densities because of the influence of a variety of
physical effects.

In a seminal paper, 
Fry \& Gazta\~naga (1993, hereafter FG) examined biasing schemes in which the
galaxy density is an arbitrary function of the local mass density.
FG expand the biasing function in a Taylor series, and they show that
if the cumulants of the mass density field exhibit hierarchical relations,
then the cumulants of the locally biased galaxy density field also
exhibit hierarchical relations in the limit that  
$\langle \delta^2 \rangle \ll 1$.
FG examined only one-point distribution functions, but their approach can
be generalized to deal with correlation functions at non-zero
separation [see, e.g., Fry (1994)
for a discussion of the three-point function].

The arguments in \S\S 2 and 3 below extend the FG results in two ways.
First, we show that if the mass clustering follows a hierarchical
pattern, then local bias multiplies the autocorrelation
function by a constant factor on large scales (where $\xi \ll 1$),
even if the bias is applied on a scale where the density field is nonlinear.
We then show that this result carries over to stochastic local biasing,
in which the galaxy density is a random variable whose mean value is
a local function of the matter density.
In \S 4 we show how our results for the galaxy autocorrelation
function translate into results for the power spectrum.
We summarize our conclusions in \S 5.

\section{Deterministic Local Bias}

A general form of deterministic local bias relates
the density fluctuation field of the galaxies, $\delta_g$,
to the density fluctuation field of the matter, $\delta$,
at the same point $\xvec$ through an arbitrary function $f$:
\begin{equation}
\label{local}
\delta_g(\xvec) = f[\delta(\xvec)].
\end{equation}
[We use quantities without subscripts, such as $\delta$
and $\xi$, to refer to the underlying matter distribution,
and subscripted quantities like $\delta_g$ and $\xi_g$ to 
refer to the biased distribution of galaxies.]
Although equation (\ref{local}) represents the most general
form of local bias in which $\delta_g$ is a function only
of $\delta$, one could imagine more general local functions
in which $\delta_g$ is also a function of, for example,
the local velocity field or
derivatives of the local gravitational potential.
Implicit in equation~(\ref{local}) is a smoothing scale
on which the continuous fields $\delta(\xvec)$ and $\delta_g(\xvec)$ are
defined.  Physically, this scale indicates the range over which
the environment directly influences the efficiency of galaxy formation.
In a random field with significant long wavelength power, the local density
contrast is itself correlated with the density contrast on larger scales,
and it is this correlation that allows a local transformation
to amplify $\xi(r)$ by a constant factor on large scales
(Kaiser 1984).  However, the Bower et al.\ (1993) model for
scale-dependent bias effectively incorporates an ``influence''
scale of 10's of Mpc, implying that a forming galaxy is
``aware'' of the physical conditions far away.

Coles (1993) and FG have demonstrated a number of important
properties of biasing models defined by equation~(\ref{local}).
For the case where $\delta$ is a Gaussian field, Coles (1993)
shows that $\xi_g(r) \propto \xi(r)$ on all scales where $\xi(r) \ll 1$
for almost any choice of the function $f$.  In other words,
arbitrary local biasing of a Gaussian density field does not
alter the shape of the autocorrelation function on large scales.
Coles also notes
that his argument fails for some simple, albeit physically
unlikely functions, such as $\delta_g = \delta^2 - \langle \delta^2 \rangle $.
FG expand the function $f$ in a Taylor series:
\begin{equation}
\label{Taylor}
f(\delta) = \sum_{k=0}^\infty {b_k \over k!} \delta^k,
\end{equation}
where $b_0$ is chosen to give $\langle \delta_g \rangle = 0$.
They then derive the cumulants of $\delta_g$ in terms of the cumulants
of $\delta$ and the biasing coefficients $b_k$, in the limit
that $\langle \delta^2 \rangle \ll 1$.
In this limit, it is obvious from equation~(\ref{Taylor})
that the leading-order effect on the variance is $\sigma_g^2 = b_1^2\sigma^2$;
in FG's notation 
\begin{equation}
\label{FG}
\overline{\xi}_{g,2} = b_1^2 \overline{\xi}_2 + {\cal O}(\overline{\xi}_2^2),
\end{equation}
where $\overline{\xi_2} = \langle \delta^2 \rangle = \sigma^2$
(see FG, equation~[9]).

With the FG expansion (equation \ref{Taylor}), the galaxy autocorrelation
function can be written
\begin{eqnarray}
\xi_g(\xvec_1, \xvec_2) &=& \langle \delta_g(\xvec_1) \delta_g(\xvec_2)\rangle, \\
\label{expand}
 &=& \sum_{j,k = 0}^\infty {b_j b_k \over j! k!} \langle \delta(\xvec_1)^j
 \delta(\xvec_2)^k \rangle .
\end{eqnarray}
If the smoothing scale on which $\delta(\xvec)$ is defined is large enough,
then $\langle \delta^2 \rangle \ll 1$, and only the $j=k=1$ term survives,
implying that
\begin{equation}
\label{linear}
\xi_g(\xvec_1,\xvec_2) = b_1^2 \xi(\xvec_1,\xvec_2) + {\cal O}(\xi^2).
\end{equation}
In other words, if there is a deterministic local relation between
galaxy density and mass density on some scale in the linear regime,
then the autocorrelation function in the linear regime
is multiplied by a scale-independent factor $b_1^2$,
where $b_1$ is the first derivative of the local bias function $f(\delta)$
evaluated at $\delta=0$.
This argument is a trivial extension of the one-point argument for
the variance given by FG, analogous to Fry's (1994) extension of
the FG skewness result to the three-point correlation function.

What if the density field is nonlinear on the scale where local
bias operates, so that $\langle \delta^2 \rangle \ga 1$?
This situation is physically plausible, and we are no longer
free to discard the higher-order terms in the sum in equation (\ref{expand}).
We can still make progress if we introduce the
assumption that the clustering is hierarchical, i.e., the connected part of 
$\langle \delta(\xvec_1)^j \delta(\xvec_2)^k \rangle$
is given by (Peebles 1980; Fry 1984; Bernardeau 1996)
\begin{equation}
\label{hier}
\langle \delta(\xvec_1)^j \delta(\xvec_2)^k \rangle_c = C_{j,k} \langle\delta^2\rangle^{j+k-2}
\langle \delta(\xvec_1) \delta(\xvec_2) \rangle +
O(\xi^2).
\end{equation}
Although the assumption of hierarchical clustering can only be shown
to be rigorously valid in the quasilinear regime, numerical
simulations show that it holds to a fairly good approximation
even in the nonlinear regime (Colombi, Bouchet, \& Schaeffer, 1994;
Colombi, Bouchet, \& Hernquist 1996; see Suto \& Matsubara 1994
for the opposing point of view), and there are theoretical grounds
for believing that hierarchical clustering should apply
in the nonlinear regime (Davis \& Peebles 1977; Peebles 1980;
Balian \& Schaeffer 1989).
There is also support for hierarchical clustering in the observed
galaxy distribution (see, for example, Szapudi, et al. 1995),
but this is not directly relevant to our argument,
since we are interested in the dark matter clustering hierarchy, which cannot
be observed directly.
The validity of equation (\ref{hier})
for our evolved density field is the key assumption we make in this
section; it allows us to generalize equation~(\ref{linear}) to
the biasing of nonlinear fields.  Bernardeau (1996) begins with equation
(\ref{hier}) and derives a gravitationally induced ``bias", but this
differs from the arbitrary bias functions we are dealing with here.

With the hierarchical assumption, we can write
\begin{equation}
\label{expansion}
\xi_g(\xvec_1,\xvec_2) = 
\sum_{j,k = 0}^\infty {b_j b_k \over j! k!} [C_{j,k} (\sigma^2)^{j+k-2} \xi(\xvec_1,\xvec_2)
+ O(\xi^2) + \langle \delta(\xvec_1)^j \delta(\xvec_2)^k \rangle_{unconnected}].
\end{equation}
The first two terms arise from the connected part of
$\langle \delta(\xvec_1)^j \delta(\xvec_2)^k \rangle$, while the last term is
the unconnected part.  Note, however, that this unconnected part can be written
as powers of lower-order correlations, which can themselves all be expanded out
according to equation (\ref{hier}).
[There are no terms of zero-th order in $\xi$ arising from the unconnected terms of
the form $\langle \delta(\xvec_1)^j \rangle
\langle \delta(\xvec_2)^k \rangle$, because all
such terms are cancelled by other terms included in $b_0$.]
In the end, we obtain:
\begin{equation}
\label{result}
\xi_g(\xvec_1,\xvec_2) = \biggr[\sum_{jk} K_{j,k} {b_j b_k \over j! k!}
(\sigma^2)^{j+k-2} \biggr] \xi(\xvec_1,\xvec_2) + O(\xi^2),
\end{equation}
where $K_{j,k}$ is a set of constants.
Hence, we find that for $\xi \ll 1$, the quantity $ b^2 = \xi_g(r)/\xi(r)$ is
approximately constant.  
Again, we wish to emphasize that we have assumed
nothing about the linearity of the density field at the scale of biasing;
all we have assumed is the validity of equation (\ref{hier}).

Both the Coles (1993) result for Gaussian initial conditions and
the FG result (equation \ref{linear}) are special cases of our general result.
If the underlying mass density field is Gaussian, as in the
case discussed by Coles (1993),
then the density field is hierarchical in the sense that equation
(\ref{hier}) is satisified, but all of the hierarchical
coefficients $C_{j,k}$ vanish except for $C_{1,1}$, which is unity.
Then our conditions are satisfied, and $\xi_g(\xvec_1,\xvec_2) = b^2
\xi(\xvec_1,\xvec_2)$.
Formally, the Coles result holds even for the case
$\langle \delta^2  \rangle > 1$, but this is not a physically realistic
case, since the density field will be Gaussian only
for $\langle \delta^2 \rangle \ll 1$.

To obtain the FG result, we simply take
$\sigma^2 \ll 1$ in equation (\ref{expansion}).  Then the $j=k=1$
term dominates,
and we re-obtain equation (\ref{linear}).  This equation
differs from our more general result in that
if the local
bias is applied on a nonlinear scale, then all of the Taylor series
coefficients
of the bias function
contribute to determining the bias factor
on large scales, not just $b_1$.

Our argument for scale-independent bias
fails when $\xi$ becomes larger than unity, which is a good thing,
since local bias can change the shape of the autocorrelation function 
and power spectrum in this regime (\cite{weinberg95}; \cite{mann97}).
Note, however, that our argument
does hold even for the case of quadratic biasing, $f(\delta) = 
\delta^2 - \langle \delta^2\rangle$.  The reason that the Coles (1993) argument
fails in this case is
that a Gaussian density field
has no connected
higher moments, so terms linear in $\xi$ vanish.

\section{Stochastic Local Bias}

The bias model of equation~(\ref{local}) can at best be an idealization.
Even in the case where galaxies ``trace the mass,'' $f(\delta)=\delta$,
there will be Poisson fluctuations about the mean relation because
of the discrete nature of the galaxy distribution.  More generally,
we expect the probability of forming a galaxy in a given region
to depend on many factors, including the history of
accretion and mergers in the nearby environment.  Many of these factors
will be correlated with the local density, but they will not be
completely determined by it.  We can quantify our ignorance
by allowing for stochastic bias, in which the galaxy density
is a random variable which depends in some way on the underlying
matter density, but which is not completely determined by it.
Little previous work has been done on stochastic bias models,
although Pen (1997) has recently attempted to model the joint galaxy-matter
probability distribution function using
a bivariate Gaussian as a starting point.

Let us therefore assume that the galaxy density $\delta_g$
at a point $\xvec$
is a random variable $X$, which is a function of the
underlying matter density at that same point:
\begin{equation}
\label{slocal}
\delta_g(\xvec) = X[\delta(\xvec)].
\end{equation}
We again assume that $X$ includes a constant term that
gives $\langle \delta_g \rangle = 0$.  As in the case
of deterministic local bias, we assume some smoothing scale
over which $\delta_g$ and $\delta$ are defined, so that the
bias does not occur at a geometric point, but over some
small volume.
The random variable $X$ is uniquely specified
by the probability of producing a particular value of $X$
given an underlying value of $\delta$, which we
write in the standard way as $p(X|\delta)$, the probability
of $X$ given $\delta$.  Note that our assumption that the stochastic
bias is purely local is actually very restrictive.  It means,
for example, that the distribution of the random
variable $X$ is the same at every point
in space with the same $\delta$, and that there are no correlations between this
distribution at different points in space.

The probability of measuring a galaxy density $\delta_{g1}$
at the point $\xvec_1$ and a galaxy density $\delta_{g2}$ at
the point $\xvec_2$ is
\begin{equation}
\label{pgalgal}
p(\delta_{g1},\delta_{g2}) = \int p(X_1|\delta_1) p(X_2|\delta_2)
p(\delta_1,\delta_2) d\delta_1 d\delta_2 ,
\end{equation}
where we have used the ``1" and ``2" subscripts to denote
the values of $X$ and $\delta$ at the points $\xvec_1$ and $\xvec_2$
and $p(\delta_1, \delta_2)$ to denote the two-point probability distribution
of the matter density at these points.  [Note that $\delta_{g} = X$
in this equation.]
For this model, the galaxy autocorrelation function is
\begin{eqnarray}
\label{xiXX}
\xi_g(\xvec_1, \xvec_2) &=& \langle X(\xvec_1) X(\xvec_2) \rangle, \\
&=& \int X_1 X_2 P(X_1|\delta_1) P(X_2|\delta_2) p(\delta_1,\delta_2)
d\delta_1 d\delta_2 dX_1 dX_2.
\end{eqnarray}
We can perform the integration
over $X_1$ and $X_2$ to obtain
\begin{equation}
\label{stoch}
\xi_g(\xvec_1,\xvec_2) = \int \overline X(\delta_1) \overline
X(\delta_2) p(\delta_1, \delta_2) d\delta_1 d\delta_2 ,
\end{equation}
where $\overline X(\delta)$ is the mean value of $X$ for a given value
of $\delta$.  This result generalizes in a straightforward
way to all of the higher-order correlation functions.

The argument that leads from equation~(\ref{slocal}) to
equation~(\ref{stoch}) is almost trivial, but the result
is rather remarkable.  It shows that the calculation of the
correlation function for the most general possible stochastic
local biasing model can be reduced to the equivalent problem for a 
deterministic local bias, with $\overline X(\delta)$ taking the role of the
bias function.  Hence, all of the mathematical machinery developed
here and in other papers for the problem of deterministic local
bias can be used for stochastic bias.  Thus far, we have
made no assumptions about the underlying density field $\delta$.
If we now repeat our assumption from the previous section
that $\delta$ exhibits hierarchical clustering, then
we obtain the same result as in the previous section:  $\xi_g/\xi$
is constant as long as $\xi \ll 1$.
Equation (\ref{stoch}) takes a particularly simple form
if galaxies trace the mass on average, $\overline X(\delta) = \delta$.
In this case, we obtain simply
$\xi_g(r) = \xi(r)$.  This result tells us that the random
fluctuations about the mean density make no difference
in the final autocorrelation function.

These results may seem counterintuitive, since stochastic
bias ought to introduce some sort of increased ``scatter" in
the final density distribution, and it certainly increases the final
rms density fluctuation.
One must remember, however, that
$\xi_g$ represents a volume-averaged correlation function, within
which all of the random fluctuations have been averaged out.
What does change for the case of stochastic bias are the random
fluctuations relative to $\xi_g$.  The variance of the autocorrelation
function at some fixed separation
is
\begin{equation}
\label{var1}
\sigma_\xi^2 = \int (\delta_{g1} \delta_{g2})^2 p(\delta_{g1}, \delta_{g2}) d\delta_{g1}
d\delta_{g2} - \biggl[\int \delta_{g1} \delta_{g2} p(\delta_{g1}, \delta_{g2}) d\delta_{g1}
d\delta_{g2} \biggr]^2.
\end{equation}
In terms of our stochastic bias function $X(\delta)$, this becomes
\begin{equation}
\label{sigma}
\sigma_\xi^2 = \int \overline {X_1(\delta_1)^2} ~ \overline
{X_2(\delta_2)^2} p(\delta_1, \delta_2) d\delta_1 d\delta_2
- \biggl[\int \overline X_1(\delta_1) \overline
X_2(\delta_2) p(\delta_1, \delta_2) d\delta_1 d\delta_2\biggr]^2.
\end{equation}
To illustrate the way in which $\sigma_\xi^2$ is increased,
we consider again the simple class of models in which 
$\overline X(\delta)=\delta$, and we we use equation
(\ref{sigma}) to calculate the difference
between $\sigma_\xi^2$ for the stochastic case and 
$\sigma_\xi^2$ for the deterministic case $\delta_g=\delta$:
\begin{equation}
\label{sigmadif}
\sigma_\xi^2({\rm stochastic}) - \sigma_\xi^2({\rm deterministic}) =
\int \sigma_X^2(\delta_1) \sigma_X^2(\delta_2)
p(\delta_1, \delta_2) d\delta_1 d\delta_2.
\end{equation}
Here $\sigma_X^2(\delta)$ is the variance of the distribution
of $X$ for a given value of $\delta$,
\begin{equation}
\sigma_X^2(\delta) = \overline {X(\delta)^2} - \overline X(\delta)^2.
\end{equation}
Since $\sigma_X^2(\delta)$ is positive, this result shows that
randomness in the bias function increases
the fluctuations about the mean value of $\xi_g$.
Conceptually speaking, equations (\ref{var1})-(\ref{sigmadif})
presume that one estimates $\xi_g(r)$ from many different pairs
of positions with spatial separation $r$ (or from a single
pair of positions in an ensemble of universes) and
computes the variance $\sigma_\xi^2$ of these estimates.
In practice, one must average over a large number of
position-pairs in order to get an estimate of $\xi_g$ that
is at all useful, but stochastic biasing will still act to increase
the variance in estimates of $\xi_g$ from one volume of the universe
to another.
These fluctuations, which can be measured in large redshift surveys,
encode information about the degree of stochasticity
in the galaxy formation process at fixed local mass density.
(Of course, even in the absence of stochastic bias, the variance
in $\xi_g$ is nonzero).

The rms fluctuation of a smoothed field can be written as
an integral over $\xi(r)$.
The conclusion that $\xi_g(r)=\xi(r)$ for $\overline X(\delta) = \delta$ 
at first seems to contradict the obvious fact that
stochasticity will increase the rms fluctuations smoothed
on any length scale.  However, these two results are not contradictory.
Recall that we assumed that the initial density field is smoothed
over some scale $R_s$, and that local bias operates over this
same smoothing scale.  Our assumption that the distribution
of $X$ is uncorrelated at different points is invalid for separations
less than $R_s$, which means that
equation (\ref{pgalgal}) also fails on such short separations.
This is most obvious for the case of zero separation.  If we measure
the density at a single point $\xvec_1$, then the product
of probabilities $P(X_1|\delta_1) P(X_2|\delta_2)$
in equation (\ref{xiXX}) must be replaced by the single
probability $P(X_1|\delta_1)$, and equation (\ref{xiXX}) becomes
\begin{eqnarray}
\xi_g(\xvec_1, \xvec_1) &=& \langle X(\xvec_1) X(\xvec_1) \rangle, \\
&=& \int X_1 X_1 P(X_1|\delta_1)  p(\delta_1)
d\delta_1  dX_1 .
\end{eqnarray}
Integrating over $X_1$ gives
\begin{equation}
\xi_g(\xvec_1,\xvec_1) = \int \overline{X(\delta)^2} p(\delta) d\delta.
\end{equation}
For the deterministic case where $\delta_g = \delta$, the corresponding
quantity is
\begin{equation}
\xi_g(\xvec_1,\xvec_1) = \int \delta^2 p(\delta) d\delta.
\end{equation}
For the special case where $\overline{X}(\delta) = \delta$, we
have:
\begin{equation}
\xi_g({\rm stochastic}) - \xi_g({\rm deterministic})
= \int [\overline{X(\delta)^2} - \overline{X}(\delta)^2] p(\delta) d\delta.
\end{equation}
But $\overline{X(\delta)^2} - \overline{X}(\delta)^2 >0$ for all values
of $\delta$, so $\xi_g({\rm stochastic}) - \xi_g({\rm deterministic}) >0$.
Thus, stochastic bias increases the rms fluctuations, but the entire
effect is due to the change in $\xi_g(r)$ at separations smaller than
our initial smoothing length; at these length scales our arguments
regarding the effects of stochastic bias on the autocorrelation function do not apply.

\section{The Power Spectrum}

Although we have focused so far on the autocorrelation function,
many observational studies of large scale structure use its Fourier
transform, the power spectrum, to quantify clustering on the largest
scales.  The Mann et al.\ (1997) numerical study of local biasing
focuses mainly on the power spectrum.
The mass power spectrum $P(k)$ is related
to the mass autocorrelation function $\xi(r)$ by
\begin{equation}
\label{power1}
P(k) = 4\pi \int \xi(r) {\sin(kr) \over kr} r^2 dr,
\end{equation}
and the galaxy power spectrum is
\begin{equation}
\label{power2}
P_g(k) = 4\pi \int \xi_g(r) {\sin(kr) \over kr} r^2 dr.
\end{equation}
In \S 2, we showed that deterministic local bias
applied to a hierarchically clustered density field
gives $\xi_g(r) = b^2 \xi(r)$ for $\xi(r) \ll 1$, but we can
put no constraint on the bias for $\xi(r) \ga 1$.  Let
$r_0$ be a distance such that $\xi(r) \ll 1$ when $r > r_0$.
We can therefore write
\begin{equation}
\xi_g(r) = b^2 \xi(r) + \widetilde \xi(r),
\end{equation}
where $\widetilde \xi(r) = 0$ for $r > r_0$.
Substituting this equation into equations (\ref{power1})
and (\ref{power2}), we get
\begin{equation}
\label{power3}
P_g(k) = b^2P(k) + 4\pi \int_0^{r_0} \widetilde \xi(r) {\sin(kr) \over kr} r^2 dr.
\end{equation}
If we choose a fixed value of $k$ for which $kr_0 \ll 1$, then
$kr \ll 1$ over the entire range of integration in the second
term, so this integral just reduces to
$4\pi \int_0^{r_0} \widetilde \xi(r) r^2 dr$,
which is a constant, independent of $k$.
Thus, in the regime $k \ll 1/r_0$,
\begin{equation}
\label{pbias}
P_g(k) = b^2P(k) + c,
\end{equation}
where $b$ is the large scale bias factor of the autocorrelation
function and $c$ is a constant, which may be positive or negative.
This is just a  more
rigorous way of noting that the power spectrum for $k < k_0$
is dominated by the correlation function at $r > 1/k_0$,
though small scale fluctuations can add a constant offset to $P(k)$.
Equation~(\ref{pbias}) is not quite the same as a scale-independent
amplification of $P(k)$.  However, the
power spectrum estimated by Baugh \& Efstathiou (1993) from the
APM survey continues to rise out to $2\pi/k \ga 130 h^{-1}\;$Mpc,
so in realistic models the constant $c$ is likely to become
unimportant on large scales, at least until one reaches the 
turnover in the power spectrum.

\section{Conclusions}

We have shown that for a local bias function applied to a density
field with a hierarchical correlation structure, the only
effect is to rescale the autocorrelation function by
an overall bias factor $b^2$ on length scales for which
$\xi(r) \ll 1$; no change in the shape of the autocorrelation
function can be induced by such a local transformation.
For the power spectrum, for sufficiently small $k$, the
result is also a rescaling, with the possible addition of a constant
term.  Although we have assumed hierarchical clustering,
our result holds as long as
\begin{equation}
\label{weaker}
\langle \delta(\xvec_1)^j \delta(\xvec_2)^k \rangle = 
D_{j,k}\langle \delta(\xvec_1) \delta(\xvec_2) \rangle + O(\xi^2),
\end{equation}
where $D_{j,k}$ is
independent of the separation between $\xvec_1$ and $\xvec_2$.
Equation (\ref{weaker}) is
actually a slightly weaker condition than the assumption of hierarchical
clustering
(equation \ref{hier}) because the moment
on the left-hand side of equation (\ref{weaker}) is not connected.

If there is a bias between galaxies and mass (and the galaxy
morphology-density relation implies that there must be bias
for at least some kinds of galaxies), then the
physics that causes it may well be complex.  However, our
stochastic biasing result implies that all environmental
effects on the efficiency of galaxy formation influence 
$\xi_g(r)$ only to the extent that they are correlated with
the mass density itself, and if these effects are local,
then they still will not change the shape of the autocorrelation
function on scales in the linear regime.  
Cen \& Ostriker (1992, figure 4) presented a first attempt
to calculate the full distribution function $P(\delta_g|\delta)$,
using a hydrodynamic 
cosmological simulation of the standard cold dark matter model.
We can expect substantial progress from this {\it a priori}
approach to biased galaxy formation over the next few years,
since advances in computer power and algorithms now allow
simulations of much higher dynamic range and permit broader explorations
of cosmological parameter space.  However, our results
imply that all of these calculations should produce galaxy
populations with $\xi_g(r) \propto \xi(r)$ on large scales.
Only a biasing mechanism that coherently modulates galaxy
luminosities on scales larger than those over which the matter
actually moves, e.g., suppression or enhancement of star formation
by quasar radiation (\cite{babul91}; \cite{bower93}),
can rescue a cosmological model that predicts the wrong shape
for $\xi(r)$ on the scales where $\xi(r) \ll 1$.  Since a physical
mechanism of this sort would surely have a different impact
on galaxies of different luminosities and morphological types,
the giant redshift surveys becoming available in the next few
years will allow us to test whether non-local biasing occurred
in the real universe by comparing the large-scale correlation
functions of different classes of galaxies.

\newpage

{\bf Acknowledgments}

R.J.S. and D.H.W. were supported in part by the NASA Astrophysical
Theory Program through grant NAG 5-3111 at Ohio State.
R.J.S. was supported in part by NASA (NAG 5-2788) at Fermilab and
by the DOE at Fermilab and at Ohio State (DE-FG02-91ER40690).

\end{document}